\documentclass{basi}
\usepackage{epsfig}

\begin{document}
\title[Twin Telescope observations of the Sun]{Twin Telescope observations of the Sun at Kodaikanal Observatory} 
\author[Jagdev Singh and B. Ravindra]%
       {Jagdev Singh\thanks{e-mail: jsingh@iiap.res.in} and B. Ravindra \\ 
        Indian Institute of Astrophysics, Koramangala, Bangalore 560 034 \\ 
       }

\pubyear{2012}
\volume{40}
\status{in press}
\date{Received 2011 December 15; accepted 2012 January 30}

\maketitle
\label{firstpage}
\begin{abstract}
We report the design, fabrication and installation of a `Twin Telescope'
at Kodaikanal Observatory intended to augment the ongoing synoptic 
observations of the Sun that has been carried out since 1904.
The telescope uses a 15 cm objective capable of taking Ca-K line filtergrams
and photoheliograms in continuum of the full disk of 
the Sun simultaneously, at a frequency of 0.1 Hz using 2k$\times$2k 
format CCD cameras. The telescope has been in operation since February 2008 and 
images are being obtained at a cadence of 5 min during normal 
observing periods. In case of 
solar activity, images of the active regions can be taken 
at a frequency of 1 Hz by restricting the field of view and spatial 
resolution. In this paper, we describe the telescope, instruments, image acquisition,
data calibration and image processing. 
We also discussed a method of determining the 
network element and plage area index. The preliminary results show that while the 
network element covers 
about 30\% of the disk, the percentage of the network element area index varies 
marginally with the seeing conditions during the day. 
\end{abstract}

\begin{keywords}
telescopes -- Sun: faculae, plages -- Sun: general -- Sun: activity  
\end{keywords}

\section{Introduction}
\label{sec:intro}
It is now well known that Sun's ever-changing magnetic field affects the solar
irradiance variability on long time scales (Harvey \& White 1999) and space weather 
on short time scales (Srivastava et~al. 2009). Long-term regular observations of 
the Sun are needed to understand the solar activities.
At Kodaikanal observatory we have been acquiring daily 
photoheliograms of the Sun since 1904 using a 15-cm aperture telescope.
The observations of Ca-K line were started in 1907 and H-alpha 
in 1912. A variety of scientific work have been carried out from the synoptic 
observations of the Sun. Singh \& Bappu (1981) used Ca-K line 
spectroheliograms to study the variation of network size with the phase of the solar 
cycle; Singh \& Prabhu (1985) found semi-periodic variations in the chromospheric 
rotation rate with a period of 2, 7 and 11 years. Sunspot data from broad-band 
images of the Sun obtained since 1904 have been used extensively to study the solar 
rotation and related topics by a group led by Sivaraman et~al. (1999, 2007) and 
Howard et~al. (2000). H-alpha images obtained daily since 
1912 have been used by Makarov \& Sivaraman (1989), and Makarov, Tlatov \&
Sivaraman (2001, 2003) for solar cycle studies and its evolution. 
The data have also been used to study solar 
activity (e.g., Singh \& Gupta 1995; Singh, Sakurai \& Ichimoto 2001). It is 
now understood that these uniform and contiguous data sets are 
extremely valuable for studying the  variations of magnetic field 
on the Sun over the past 100 years. To keep the data sets uniform while
acquiring data with a new telescope, it is important to continue to obtain 
data and generate a series of images from overlapping data from the earlier 
instrument and the new one.

Earlier, the images of the Sun were recorded on specialized photographic 
emulsion suitable for this purpose. With the advancement of electronic 
technology and development of faster and bigger format CCD cameras, the 
specialized films went out of production. In 1995 we started using a narrow 
band filter with the old siderostat and CCD camera of 1k$\times$1k format to take 
Ca-K line filtergrams. These data have the drawback that images rotate over 
time and have low spatial resolution compared with the earlier data obtained
with the spectroheliograph.  To overcome this limitation, we 
designed and fabricated a Twin Telescope to take Ca-K line and continuum images 
of the Sun.  This telescope has been in operation since 2008 at the
Kodaikanal Observatory and have been making images during clear skies. 

Though the new telescope installed at Kodaikanal obtains both the Ca-K and white light
images, the main focus of this paper is on Ca-K 395~nm image data obtained from 
one of the telescopes. In the following sections of the paper, we describe
the instrument used to obtain the synoptic Ca-K data sets,
the observational method and calibration techniques adopted to be make it 
useful for the scientific community. Subsequently, we present a technique 
to extract the information about the network element area and plage area 
index using the calibrated data sets. 
In the end, we summarize the instrument and the preliminary results.

\section{Instrument}

\begin{figure}
\begin{center}
\includegraphics[width=100mm]{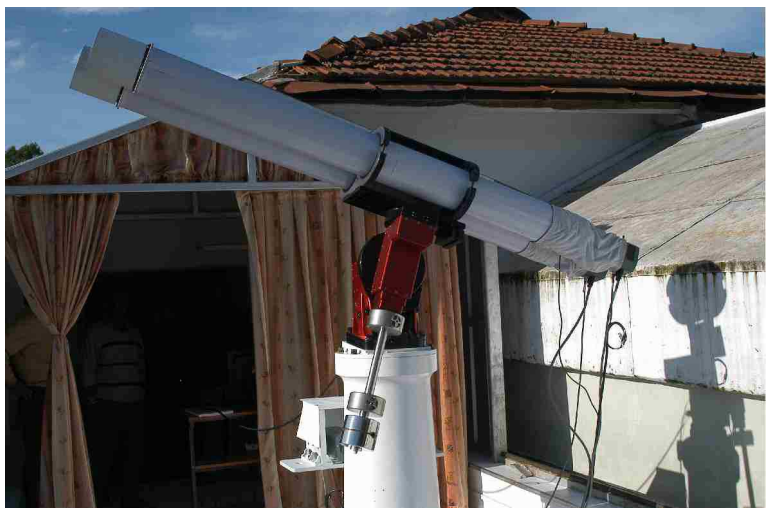}\\
\end{center}
\caption{The Twin Telescope observing the Sun at Kodaikanal Observatory.}
\label{fig:1}
\end{figure} 

The Twin Telescope consists of two tubes mounted on a single equatorial mount, 
one to obtain the white light images and the other for the Ca-K line filtergrams of the Sun 
(Fig.~\ref{fig:1}). 
Each tube is fitted with 15-cm objective lens from Zeiss to take the image of the
Sun at different wavelengths. Each lens of focal length 225~cm forms the full disk 
solar image of size 2.06~cm.
For white light images we use a heat rejection filter of size 15-cm with a pass-band 
centered at 430~nm and bandwidth of 10~nm which is kept in the entrance of the tube. 
In addition, a Mylar filter with density 
5 is kept in-front of the objective lens to cut down the incoming light intensity so as to 
reduce the heat load in the telescope tube.
The Mylar filter used in the Twin Telescope has a 
refractive index similar to that of ambient air. Index matching prevents wavefront
degradation and helps retain the desired image quality with the telescope.
On the other hand, the neutral density (ND) filter is generally used near the 
focal plane.  Thus it is not made with optical surface quality and may have some aberrations. 
The large-sized ND filters with optical surface quality have to be custom made and it is expensive.
Hence, we opted for a low cost Mylar filter to reduce the intensity and heat load in the telescope.

The telescope which takes the Ca-K image uses a heat rejection filter with a pass band 
centered at 395~nm and band-width of 10~nm kept in front of the objective lens. In addition,
a Mylar filter with density 3.8 is placed in front of the objective lens. More importantly,
a narrow band thermally controlled interference filter with a pass band centered at 393.37 nm and 
band width of 0.12 nm is kept near the focus to get the Ca-K line images. 
Thus the Ca-K line filtergrams are obtained with spectral pass band of 0.12 nm 
centered at the line. The pass band remains stable within 0.001 nm as the temperature
of the filter is maintained within 0.1$^{\circ}$C. The final image size is 2.06~cm with a 
small variation over the seasons. The image scale is 93$^{\prime\prime}$~mm$^{-1}$. Two CCD cameras, 
one for the Ca-K line and the other for the broad-band imager permit us to take 
simultaneous images of the Sun for the chromospheric and photospheric studies in case 
of active events on the Sun. 

\subsection{CCD detector and image acquisition}
The CCD cameras having scientific grade-I chip and 
2k$\times$2k pixel format with a 16-bit read out at 1~MHz made by Andor provides uniform
images with high dynamic range and high photometric accuracy. The peltier cooled 
CCD cameras are operated at --40$^{\circ}$C for low dark current and low read out noise.   
The CCD camera used in imaging has a pixel size of 13.5$\times$13.5 microns providing 
a spatial resolution of 1.25 arc sec per pixel. The 
ND filters in front of the objective lens permit us to give sufficient large exposure 
to avoid the visibility of the shutter pattern in the images of the Sun. The exposure 
time is chosen in the range between 300 ms and 1 second depending on the sky conditions.
During the clear sky the exposure time is generally 200 - 300~ms depending on the time of 
observations. In the morning, exposure time is larger due to large extinction.
Large exposure times are used during the presence of high altitude thin clouds. 
After the alignment of the telescope, the auto guider indicated 
that drift in the image is about 2-3 arcsec in one minute. This works out to be 0.015~arc-sec 
during the exposure time of 300 ms which is much less than a pixel size.
The read out noise of about 20 counts (standard deviation) determined from the dark
images which is about 0.1\% of the image count, is much less than the photon noise which is 
about 170 counts in Ca-K line images. The photon noise is about 0.6\% of the image count 
and thus dominates in the photometry. The low dark count of about 300 affects the
dynamic range of the 16-bit CCD camera marginally. 

The software program allows us to set the exposure time, rate of image 
acquisition and cadence for saving the images. The white light image of 
the Sun itself is used to center the image on the CCD detector. The developed 
software computes the edges of the solar limb and keeps it fixed at specified 
locations on the CCD cameras within a few pixels. The sequence continues to be 
recorded until we interrupt it due to the weather conditions. In principle, sequence
of the full disk images of the Sun can be obtained with an interval of 6 sec 
but we normally take the images of the Sun at an interval of 1~minute to guide the 
image and save the images at an interval of 5 min to keep the data volume at 
a manageable level. In future, we plan to save the data every 1 min or better
to examine the evolution of solar features at different wavelengths.
In case of activity on the Sun such as flares, we plan to 
obtain the images at an interval of 2 sec by restricting the FOV and binning 
the CCD chip by 2$\times$2 pixels.

In order to know the orientation of the images on the CCD we off center the image 
towards the East side on the CCD and obtain a portion of the Sun's image.
We then allow the image to drift by stopping the telescope. About 70- 80 seconds
after stopping the telescope we take another image. A combination of 
these two image gives the E-W direction of the Earth. This procedure is repeated  
once in 15 days depending on the sky conditions. We take flat field images in 
clear sky conditions, preferably once in 3 days, by keeping a diffuser in 
front of the objective lens. We also take dark images at regular intervals.

\subsection{The data calibration}
\subsubsection{Flat fielding}
We attempted a number of ways to take the flat-field images. 
First, we tried by pointing the telescope away from the Sun towards the
eastern, western, northern and southern directions and taking images of
the sky. The varying sky brightness with distance from the
Sun created a small gradient in the flat-field image. The average
of flat-field images showed a small gradient. Hence, we did not adopt 
this method. Then we
obtained the flat-field image by keeping a diffuser in front of the objective
and pointed the telescope to the disk center, but the image again
showed a small gradient in different directions. The diffuser scatters
the solar disk light in all directions in the telescope and probably
causes a gradient in the images. The outer portions of the
images of the Sun in Ca-K line do not show any arbitrary variations
in the background signal due to change in the sky brightness. 
Absence of gradients except the limb darkening
gradients in the Ca-K line images indicate that observed
large spatial scale gradients in the flat-field images are because
of the method adopted to take the flat field images. It is, therefore,
necessary to remove this effect from the flat-field images.
The flat-field correction applied to the Ca-K images of the Sun
without making the above mentioned correction to the flat-field
images caused distortions in the solar images. There are other
methods to compute the flat fielding (e.g., Kuhn, Lin \& Loranz 1991)
for extended objects. We are planning to use one of these methods in future.

The flat images are obtained with the same telescope setup by keeping a 
diffuser in front of the telescope objective. The flats are 
taken once in 3 days and sometime once in a week depending on the sky conditions. 
A set of obtained dark and flat images are averaged separately. A second degree polynomial 
is used to remove the gradient in the averaged flat field image
caused by the scattered light in the telescope due to the diffuser.  The dark 
current has been subtracted from flat field image and then normalized the flat image 
to the maximum value of dark subtracted flat. Later, all the Ca-K images are corrected 
for the flat fielding. 

\subsubsection{Image alignment}
The next step is to find the North-South direction on the
solar image for which the following steps have been adopted. 
 (1) Two solar images were obtained by stopping the telescope, one
in the eastern direction and other in the western direction as explained in Section 2.1.
(2) Wherever the Sun's disk is found in the image, those pixels were assigned a 
value of 1, so that the image has become now a binary map. (3) These images were added. 
(4) We then identified the pixels that are having maximum counts after adding the images
and made those pixel values equal to 1. The rest of the image is made zero. 
(5) Slicing the image column wise, we identified the column that has only 
one pixel and its value is 1. (6) Following the 5th step, we identified the two 
bright pixels that corresponds to the  northern and southern direction of the Earth. (7) Then the
 p-angle is calculated using the standard solarsoft routine by inserting the date and 
time of the observations. (8) The final angle for the image rotation is computed 
by considering the value of p-angle and the E-W direction of the Earth on the image
of the Sun obtained using the CCD camera. In many occasions, the obtained solar images 
are not in the center of the image window. By finding the center of the solar image 
(by centroid finding method) we calculated the required pixel shift of the solar image center
from the center of the image window. Once the image center was 
identified we then rotated the solar image so that the image North pole is 
in the upward direction.  All these data 
calibrating softwares are written in Interactive Data Language (IDL) from ITT visual. 
These calibrated images are then stored in 32 bit FITS format for scientific analysis and 
also in jpeg image format for a quick look. 

\subsubsection{Observations}

\begin{figure}
\begin{center}
\includegraphics[width=60mm]{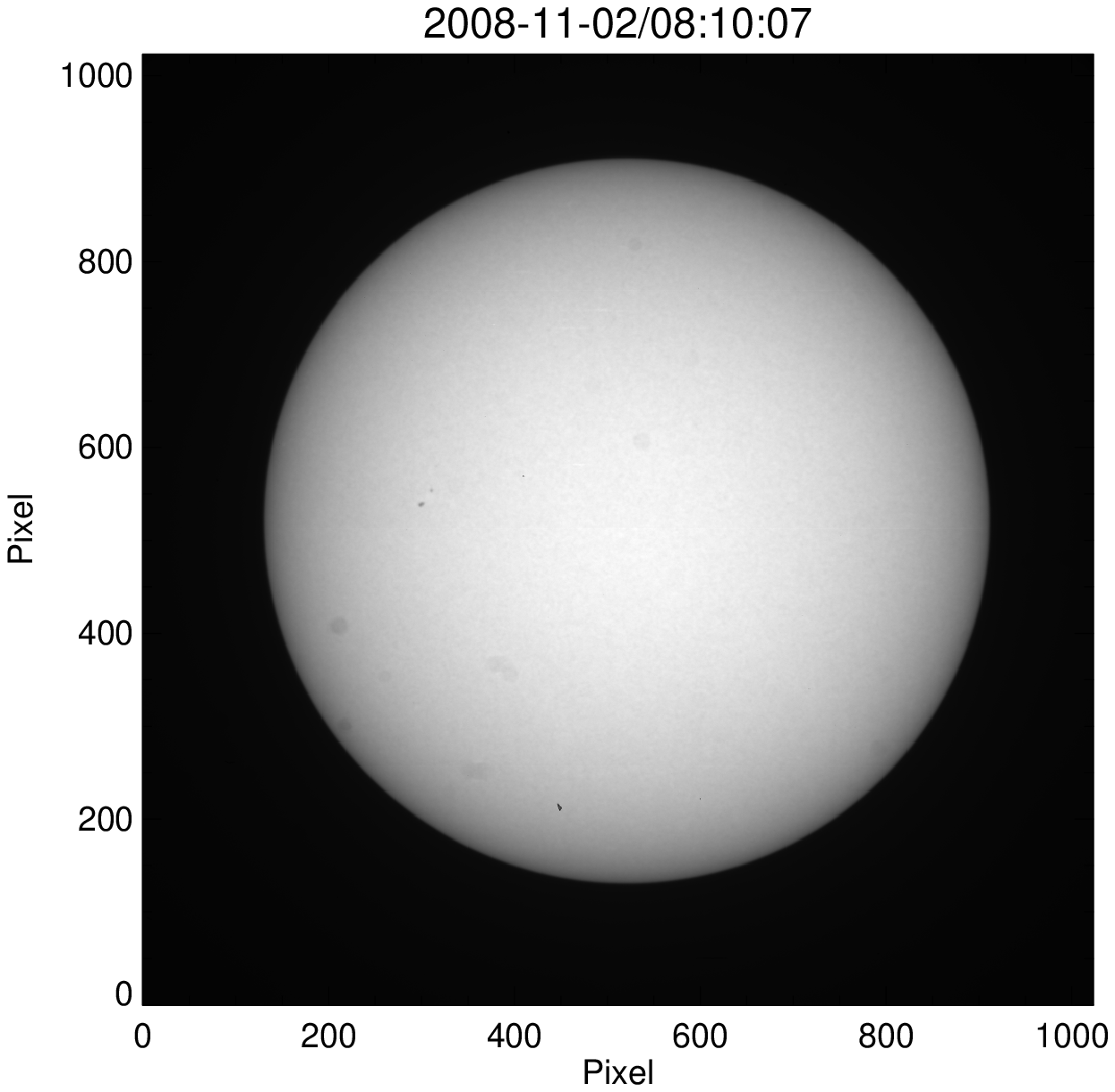}\includegraphics[width=60mm]{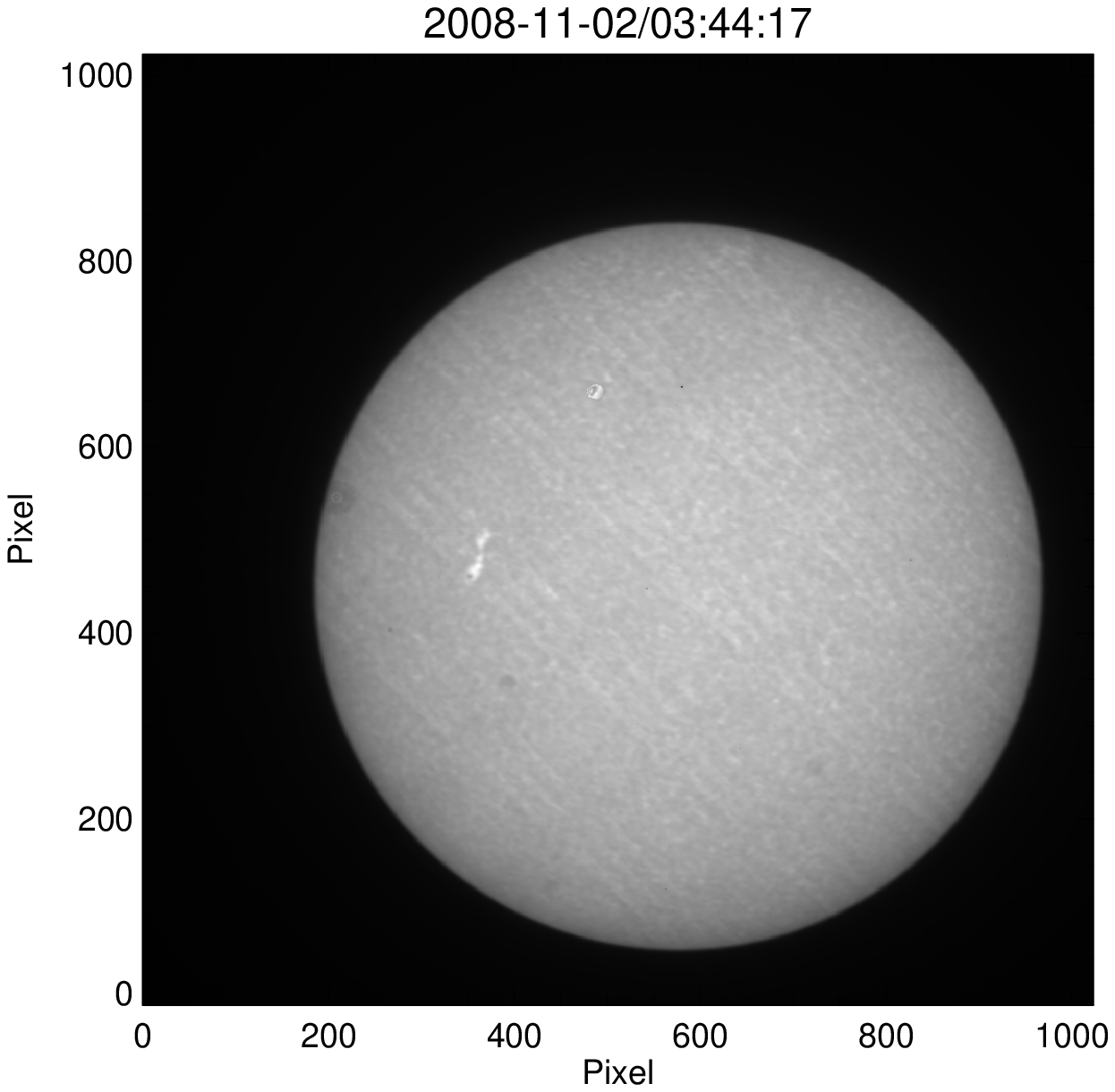} \\
\includegraphics[width=60mm]{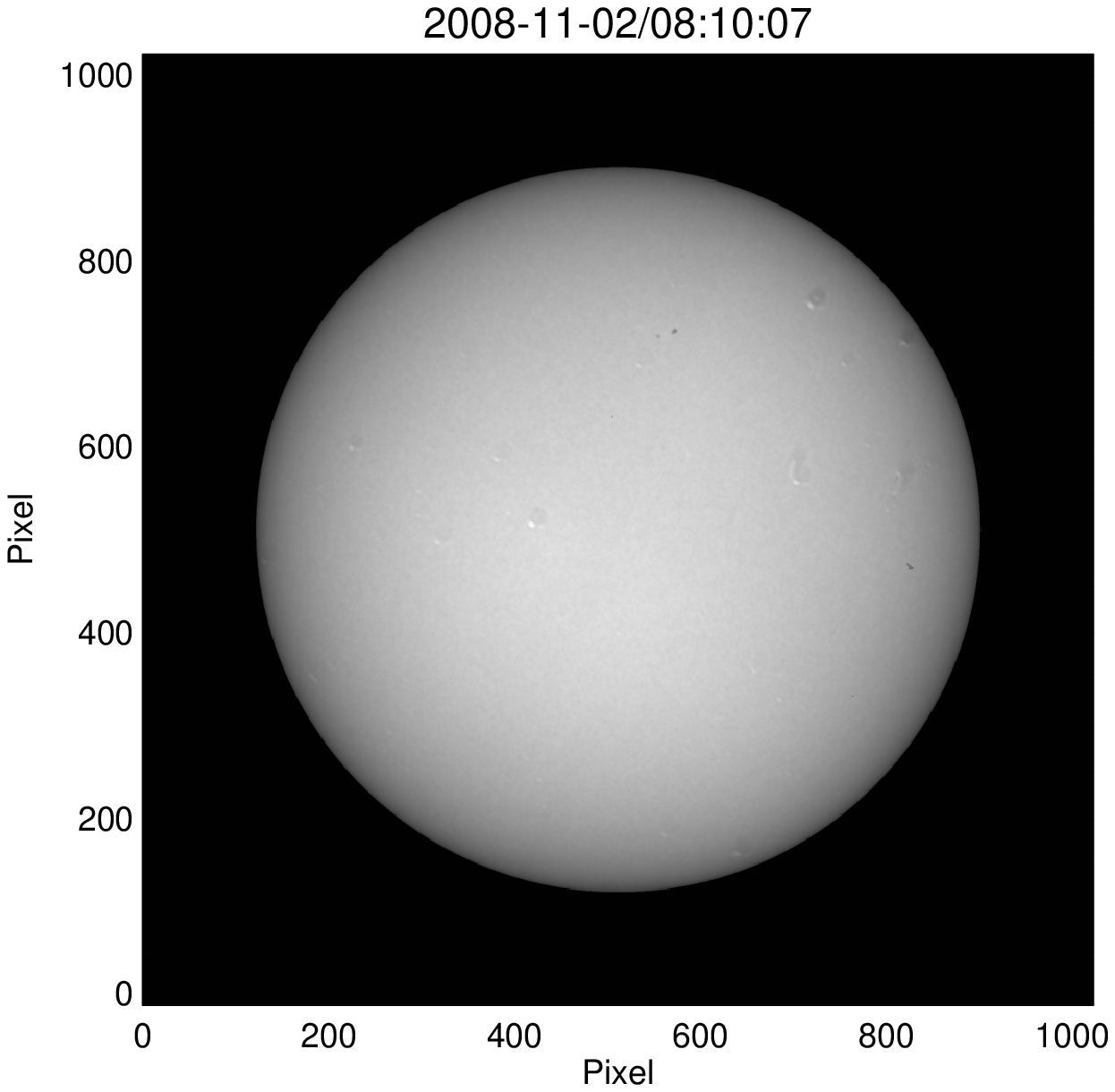}\includegraphics[width=60mm]{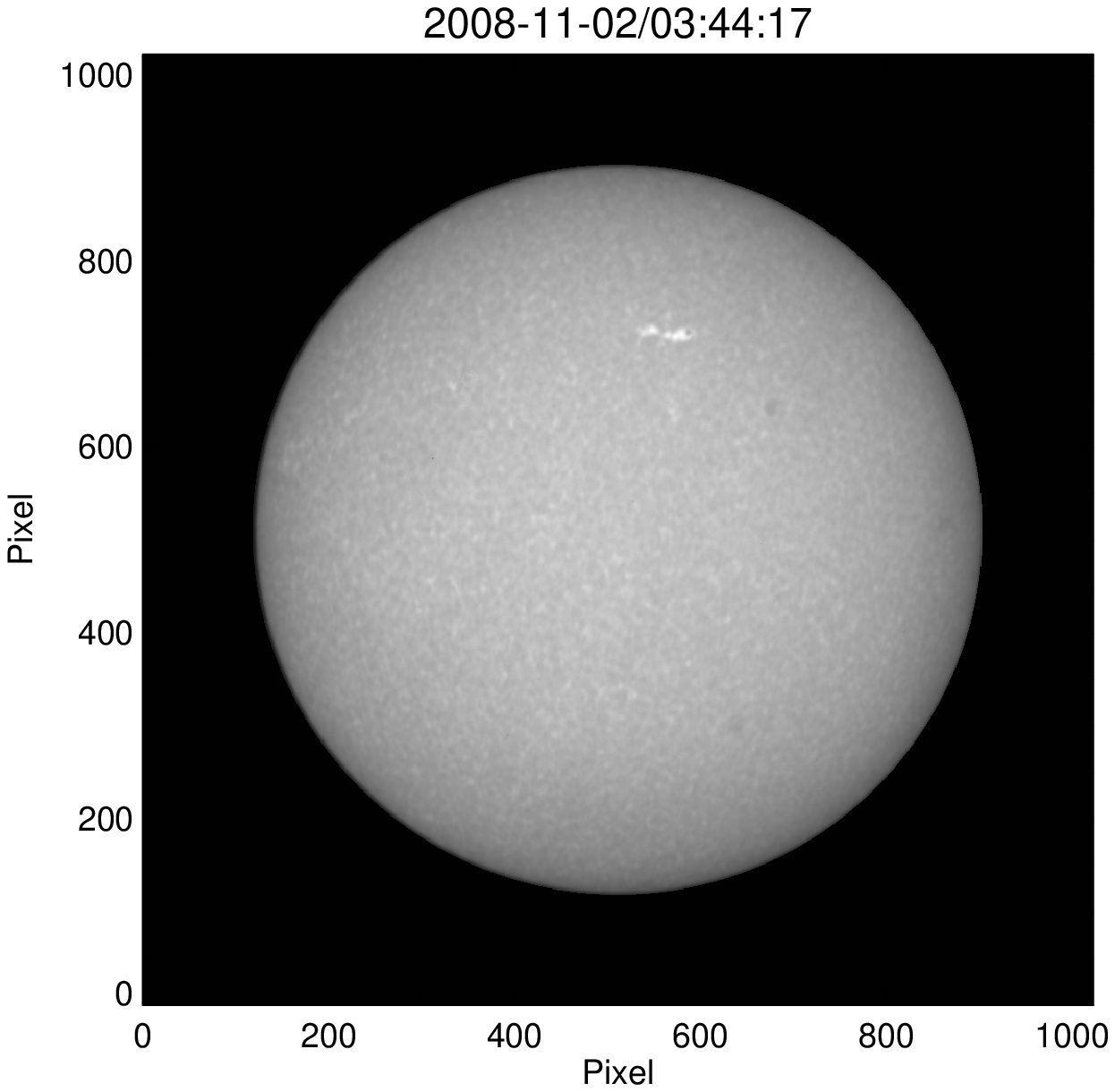}\\
\end{center}
\caption{Top left: The raw white light image. Top right: The raw Ca-K image.
Bottom left: A preliminary calibrated white light image. 
Bottom right: A Ca-K image which is flat fielded and rotated to make the North polarity up.
The date and time (in UT) of the observations are shown on the top of each image.}
\label{fig:2}
\end{figure} 

The Twin Telescope was made functional on February 23, 2008 and has been obtaining the 
Ca-K line and white light images of the Sun whenever the sky is clear. The quality of Ca-K data is 
mostly homogeneous. When the sky is clear the data is acquired once every 5 min. 
Under clear-sky conditions we are able to get upto 80 images, and upto 10 images on other days. 
The telescope is operational and continues to acquire data. 
Fig.~\ref{fig:2}(top) shows the typical raw white light (left) and Ca-K (right) 
images obtained from the 
Twin Telescope. Fig.~\ref{fig:2}(bottom) shows the corresponding images that 
are flat fielded and corrected for the North-South. In raw white-light images, dust particles
and non-solar features are clearly visible.
These features are largely diminished in the flat-fielded images.
The fringe pattern seen in raw Ca-K images has been successfully removed after the flat fielding.
However, a few dust particles 
and other features are still present but these remaining features do not affect the statistical 
studies.

\section{Image processing}
The Ca-K images show the well known limb darkening effect. In order to extract any features on the
Sun by automatic detection processes it is essential to remove the limb darkening 
gradient. In the following we describe a method to compute the limb darkening profile and
make the required correction. 

\subsection{Disk center and radius}
For many scientific data analysis it is essential to identify the center and radius of
the solar disk in the calibrated image.
This has been achieved by identifying the solar limb. Solar limb has a steep gradient 
between the solar disk and the surrounding area. We used the sobel filter to detect 
the edge of the solar limb. We then used a threshold value of five times the mean value of the 
sobel filtered image to detect the edges automatically. We assigned a value of 
one to the limb and kept the rest of the image at zero value. We first sliced the image in 
the column direction. Since the images are already rotated to make the North pole in the 
vertical direction, the column first detected on the limb with values equal 
to one is identified as the East limb of the Sun. Similarly, the last one is taken 
as the West limb. We then sliced  the image in the 
row direction. Similar as above the first detected limb with values equal to one is taken as
the South and the last is taken as the North position of the limb. To detect more
points on the limb we have generated circles centered at four corners of the image 
window. We allowed the circle to grow. Wherever it touched the Sun's limb first is 
taken as the S-E, N-E, N-W and S-W points. Using the eight detected points on the solar 
limb we used a polynomial fitting (FIT$\_$CIRCLE.PRO available in the solarsoft) to 
obtain the Sun center and radius in terms of pixels. With the adoption of this 
method we could identify the Sun center and radius.

\subsection{Removal of limb darkening}

\begin{figure}
\begin{center}
\includegraphics[width=100mm]{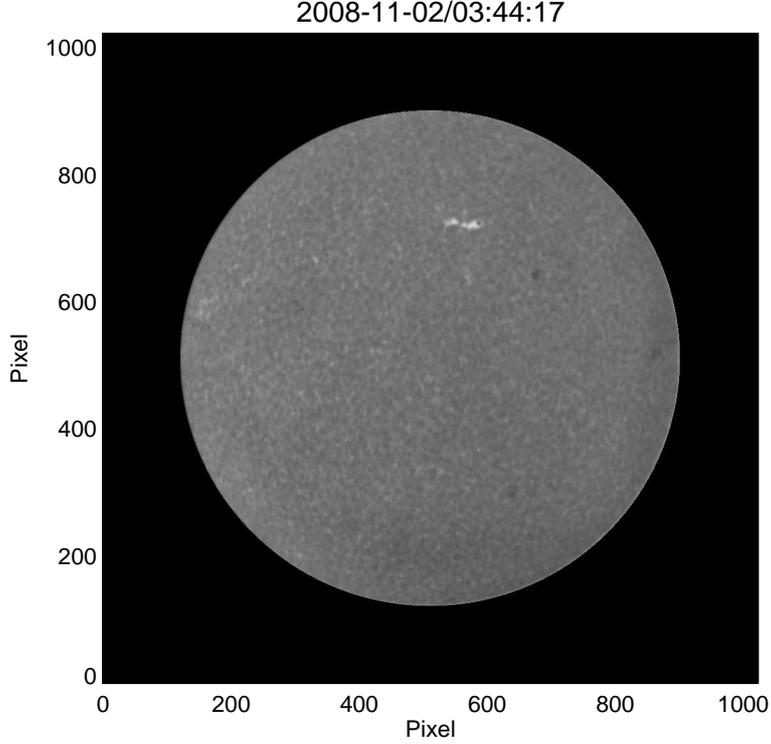}\\
\end{center}
\caption{The limb darkening corrected Ca-K image corresponding to Fig.~2 (bottom right).}
\label{fig:3}
\end{figure} 

The photospheric and chromospheric limb darkening is a gradual decrease 
in the intensity from disk center to the limb. In order to remove this systematic variations 
of intensity we followed the method described in Denker et~al. (1999). 
Following this method, we first computed a median intensity in a small box
enclosing the disk center. The data is transformed to polar co-ordinates that resulted 
in one radial profile for every 1$^{\circ}$ azimuth. A median value is obtained at each
radial position to get the average radial profile. The inner portion of the average disk 
profile is replaced with a second order polynomial fit to the average radial profile. 
Once again the average disk profile is smoothed by large smoothing kernels. The resulting 
smoothed profile is transformed to the Cartesian co-ordinate system. The pixels outside the
disk are replaced with zeros. A typical limb darkening corrected image is shown
in Fig.~\ref{fig:3}.  

\subsection{Removal of residual gradients}

\begin{figure}
\begin{center}
\includegraphics[width=100mm]{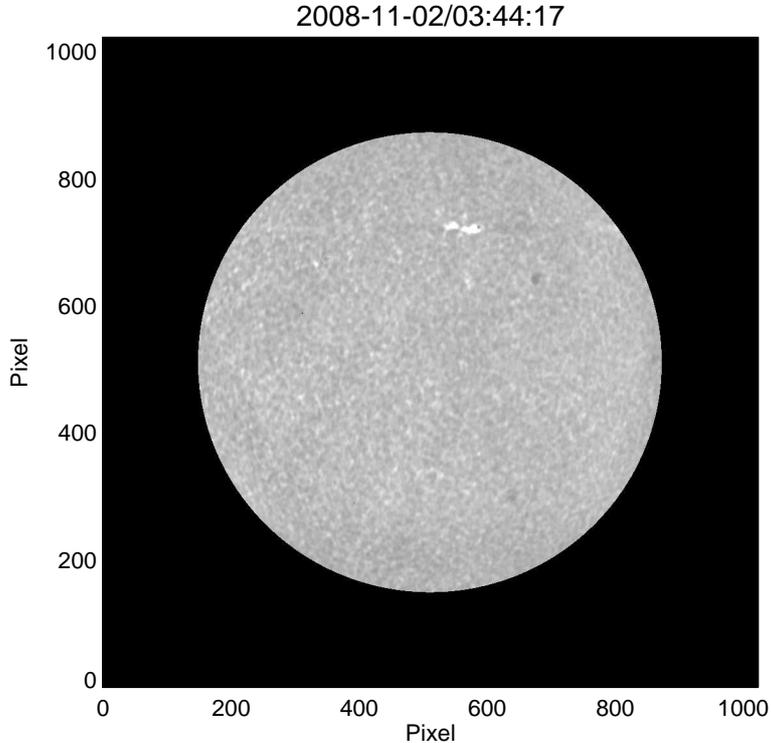}\\
\end{center}
\caption{Global fitted Ca-K image.}
\label{fig:4}
\end{figure} 

For many automatic feature identification on the solar disk it is essential to make the 
background smooth. For this purpose, we first attempted a 2-D surface fitting. Even after this, 
we observed some non-uniformity still present in the images. Later we used a 1-D polynomial
fitting to each row and column in the images. Care was taken that it does not produce any 
bias in the data but makes background more or less uniform as expected. It does not 
affect the study of properties of the solar features. We first fit a third degree polynomial
to a one dimensional, row wise sliced pixels. The resulting fitted
curve is subtracted from the original row wise sliced pixels by keeping the
mean level of the 1-D data as before. A similar fitting is done for column wise
sliced pixels. Here too we used a third degree polynomial to remove the overall trend in the images.
This procedure makes the background smooth and the resulting images (Fig.~\ref{fig:4}) show 
the network features with better contrast. Further details about this procedure can be seen in 
Singh et~al. (2012).

\section{Network element detection and area estimation}

\begin{figure}
\begin{center}
\includegraphics[width=80mm]{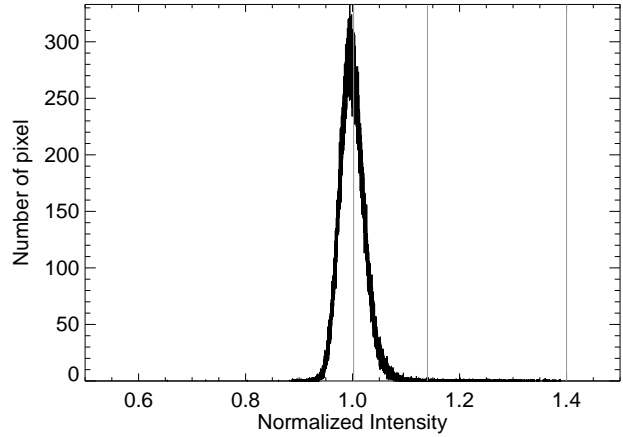}\\
\end{center}
\caption{Histogram of the Ca-K image. The three vertical lines in the plot demarcates the
intensity regions of intra network, network and plages.}
\label{fig:5}
\end{figure} 

The synoptic data is essential to study the long-term variations in the plage and network
element area index. To study the variations in area covered by these features it is essential to 
identify them automatically. One of the methods to identify the features is to use 
threshold values. The histogram of intensity values helps in deciding the threshold values.
Fig.~\ref{fig:5} shows the histogram of the intensity distribution in Ca-K images.
The intensity values have been normalized to the background intensity.
In the plot, the region between the first and second vertical lines represents the network. 
Similarly, the region between the second and third vertical lines represents the plage regions. 
The intensity before the first vertical line corresponds to the intra network regions. 
In a distribution curve of various intensities in a Ca-K image,
the peak value represents the mean value of the background intensity. 
If we normalize the whole image with respect to their mean value 
then the back-ground intensity value becomes about one.  Then with respect to this normalized 
value the network elements and plages will have higher values.

\begin{figure}
\begin{center}
\includegraphics[width=80mm]{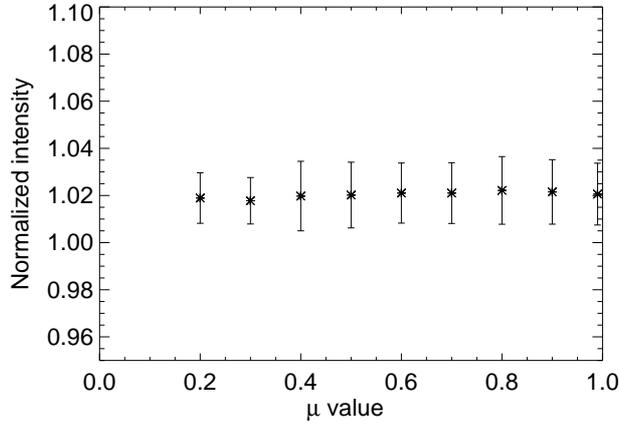}\\
\end{center}
\caption{The intensity contrast of the network elements extracted from the region of ring 
with a width of $\mu$=0.1. The vertical bars show the range of contrast values over the mean.}
\label{fig:6}
\end{figure} 

\subsection{Center to limb variation of network element contrast}

Once the limb darkening correction is made and globally fitted to make the background 
uniform, the intensity contrast of the network elements become uniform from center to the
limb. Fig.~\ref{fig:6} shows the center to limb variation of the contrast values of
the network elements.  The intensity contrast has been estimated at each
$\mu$(=cos$\theta$) value and averaged over a width of $\mu=0.1$ and has been shown in 
the plot. The angle $\theta$ is computed from the center of the disk towards the edge of the 
disk. The value of angle $\theta$ is
zero in the center and it is equal to 90$^{\circ}$ at the limb. We avoid the limb as 
the value of cos$\theta$ (=$\mu$) will approach zero. The vertical bars represent 
the variations in the intensity contrast in a ring of $\mu$ = 0.1. From the plot it is clear that
the mean value of intensity contrast of the network elements is 1.02 and it varies from 1.006
to 1.035. Hence, the use of intensity threshold values with lower limit of 1.006 and upper 
limit of 1.04 should pickup all the network elements present on the disk. However in this 
study, we consider the enhanced network elements also as a part of the quiet network. The intensity
contrast value of enhanced network is a little larger than 1.04 and  we chose a upper threshold value 
of 1.14 to detect the quiet and enhanced network elements.

\begin{figure}
\begin{center}
\includegraphics[width=60mm]{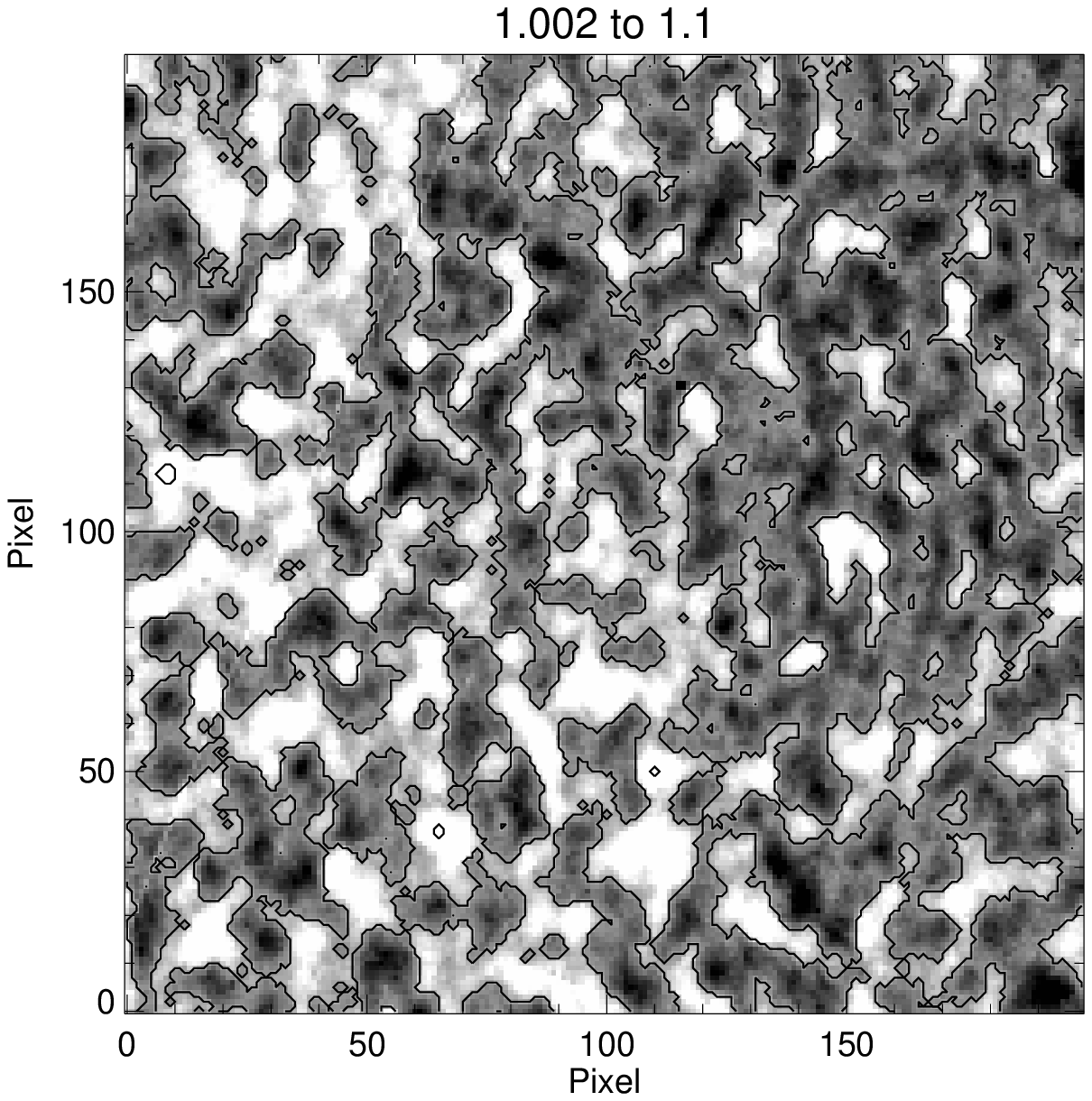}\includegraphics[width=60mm]{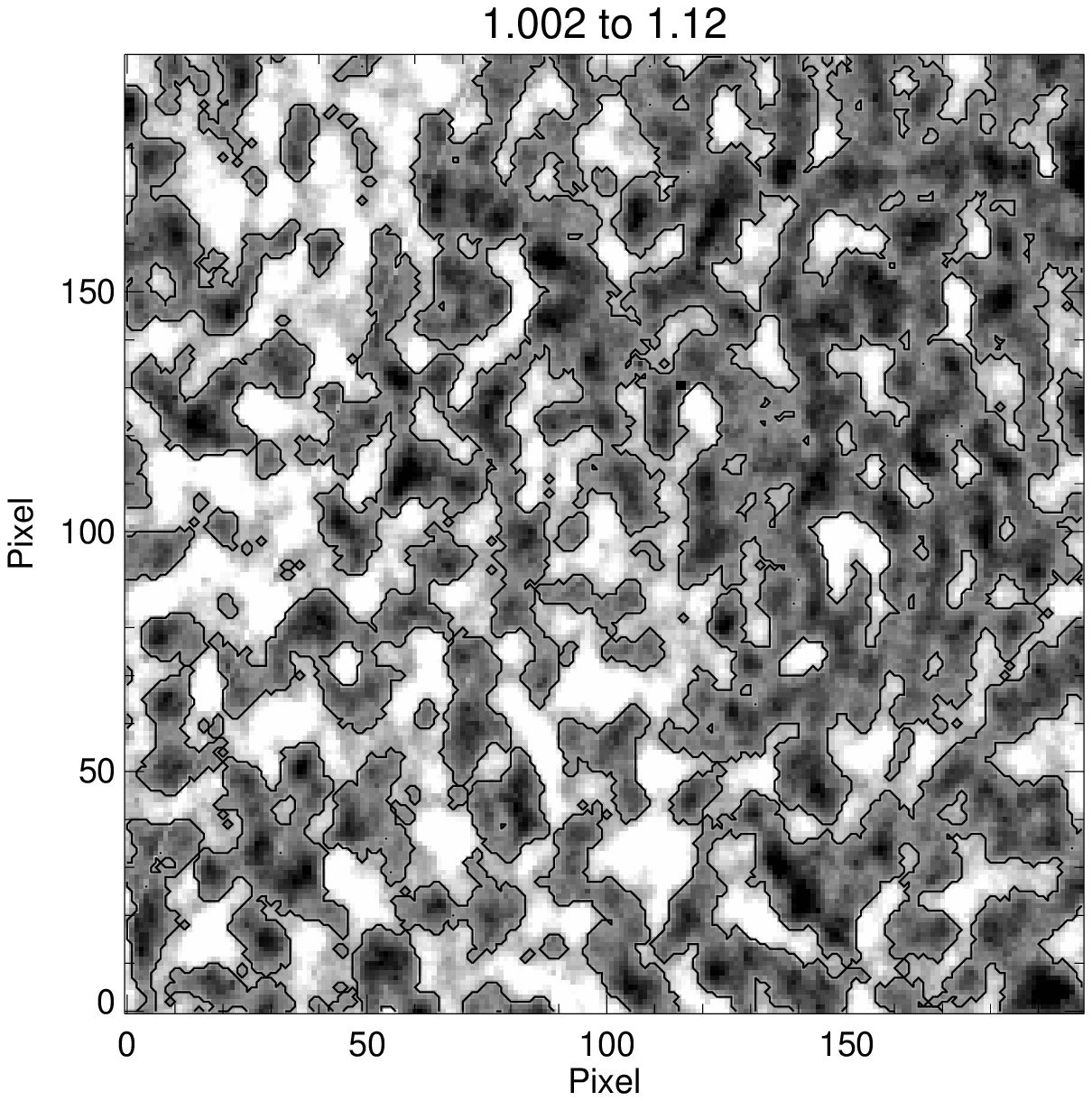} \\
\includegraphics[width=60mm]{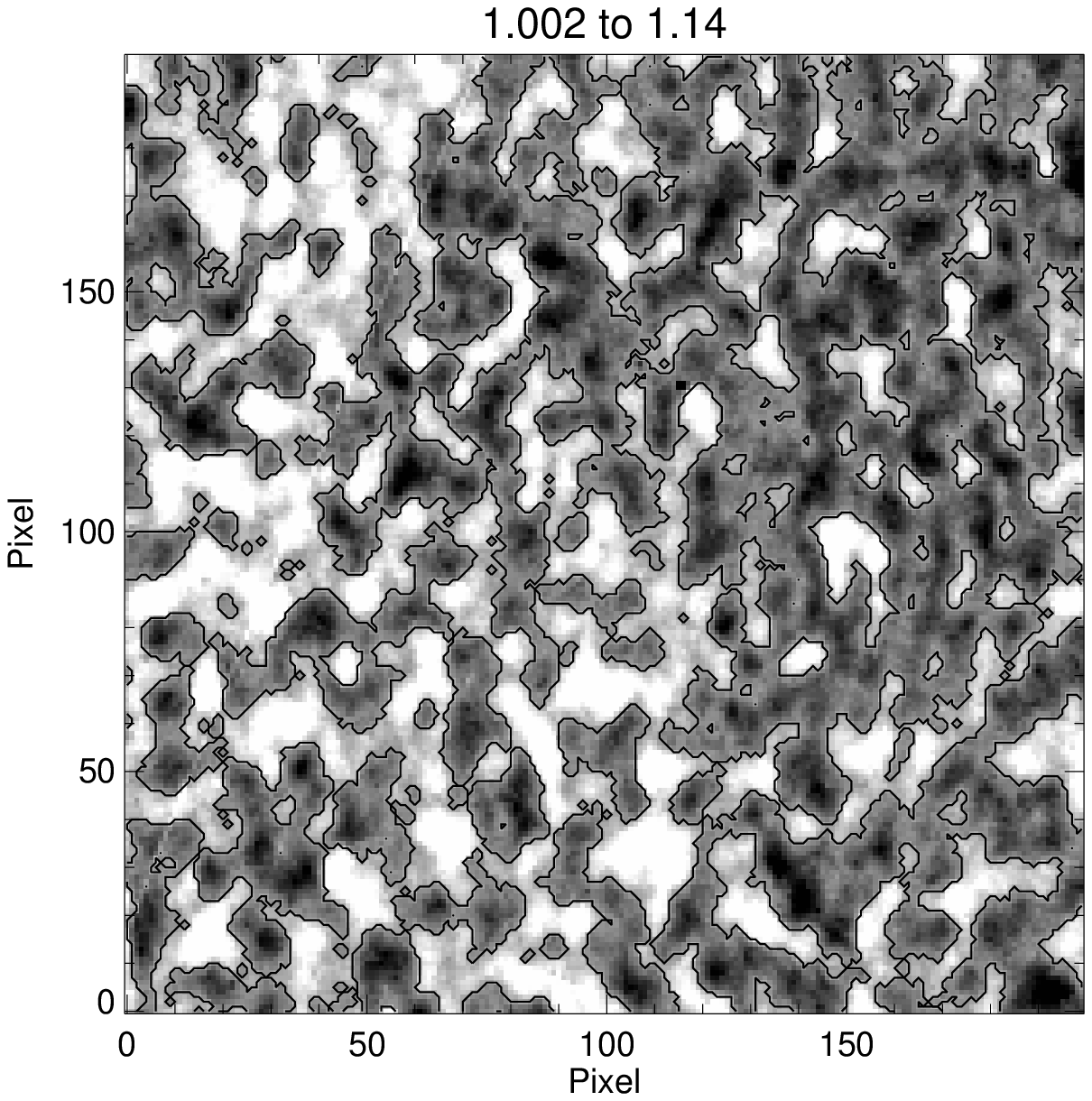}\includegraphics[width=60mm]{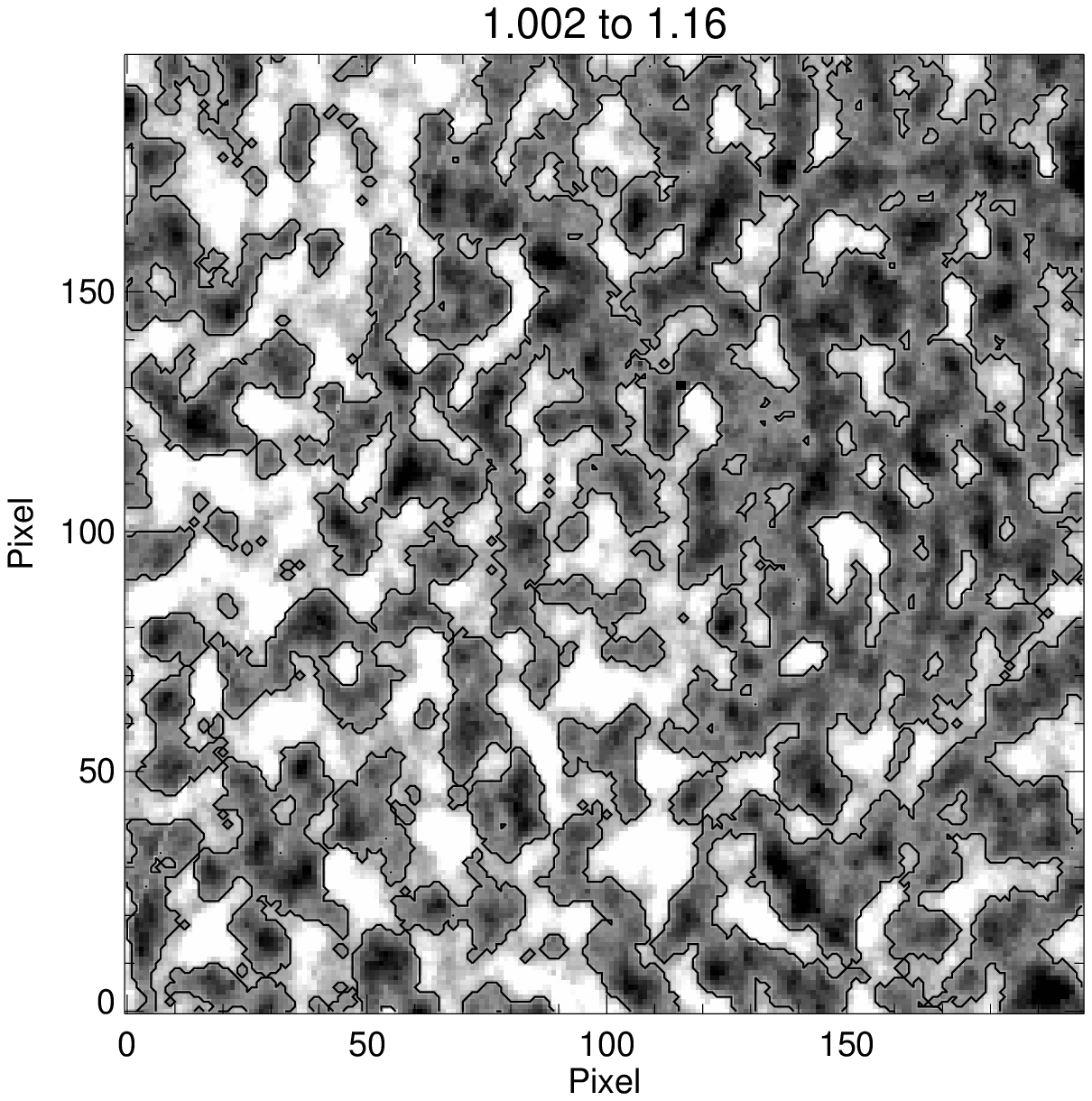} \\
\includegraphics[width=60mm]{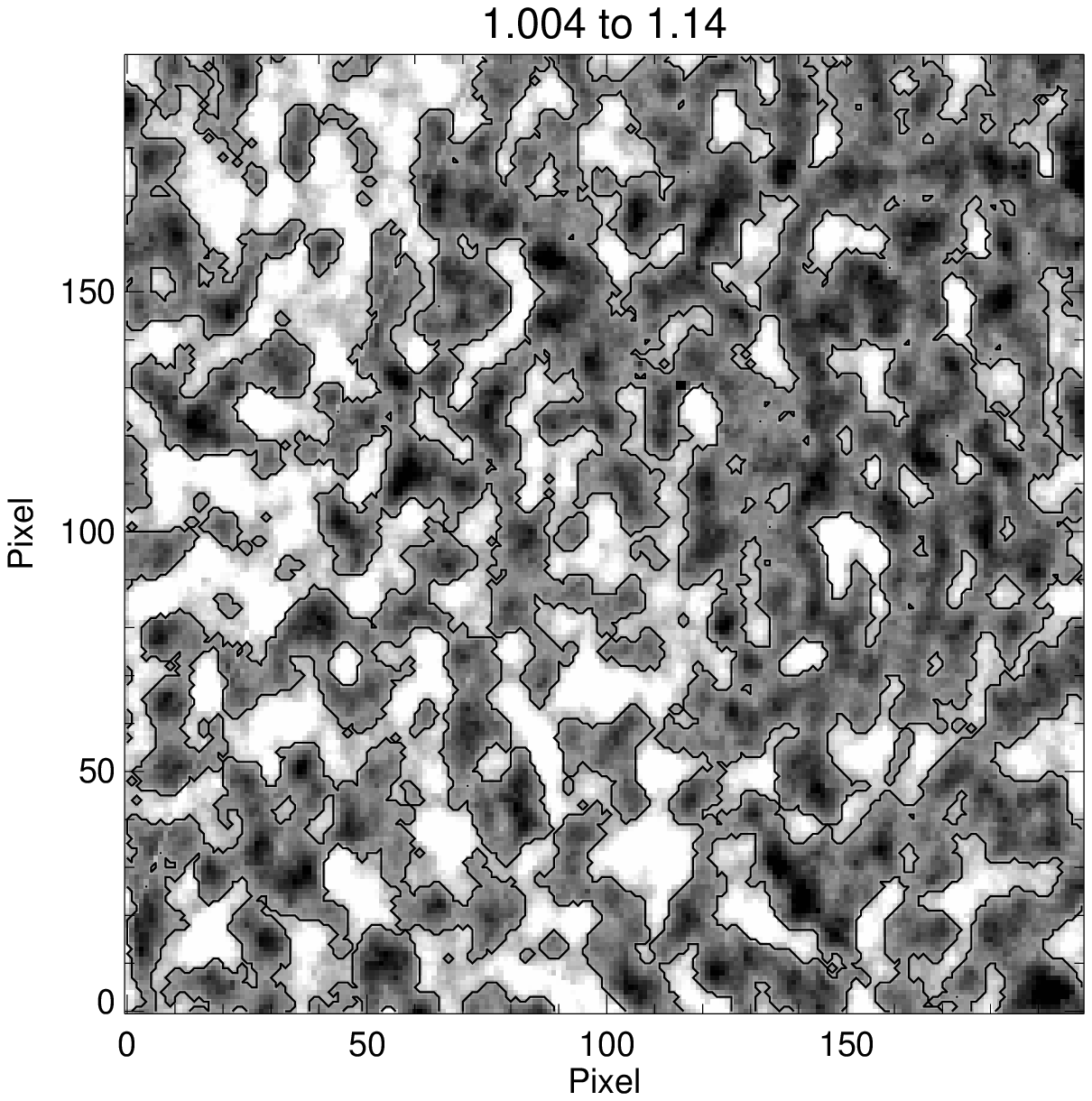}\includegraphics[width=60mm]{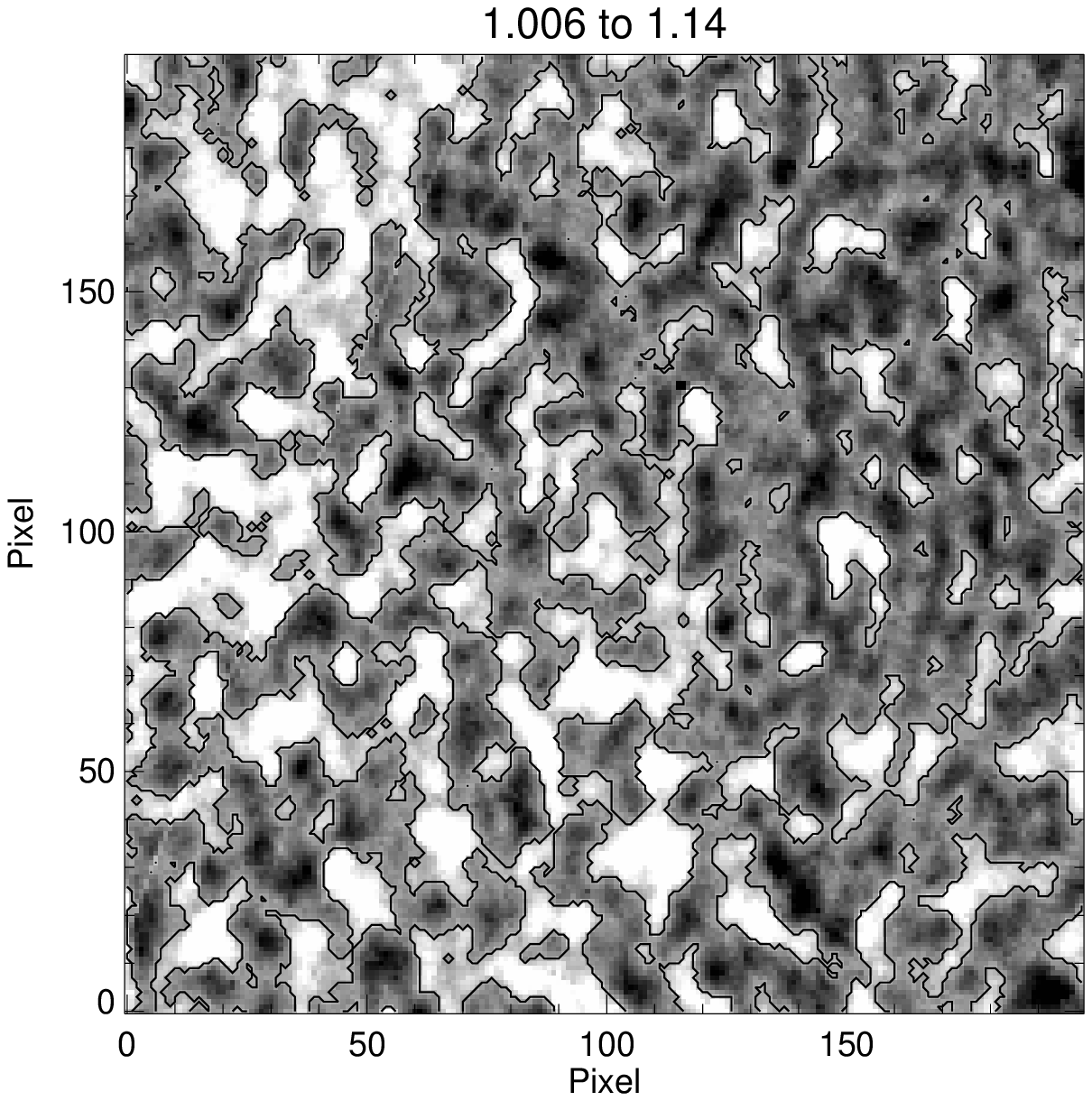} \\
\end{center}
\caption{Contour map of detected network elements for different threshold values are overlaid
upon the limb darkening removed image. The lower and upper threshold values of image contrast
are shown on the top of each contour map.}
\label{fig:7}
\end{figure} 

\begin{figure}
\begin{center}
\includegraphics[width=60mm]{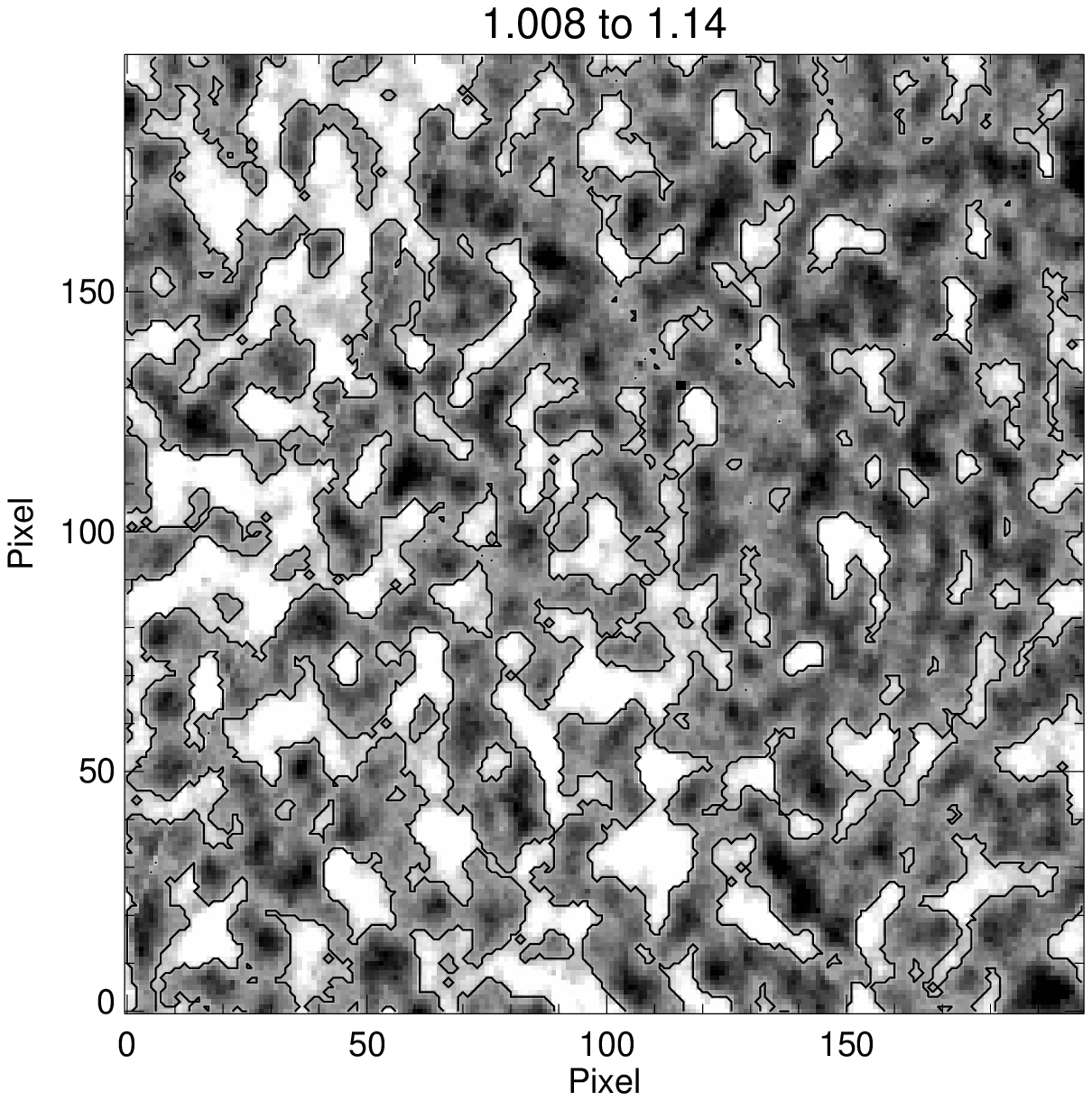}\includegraphics[width=60mm]{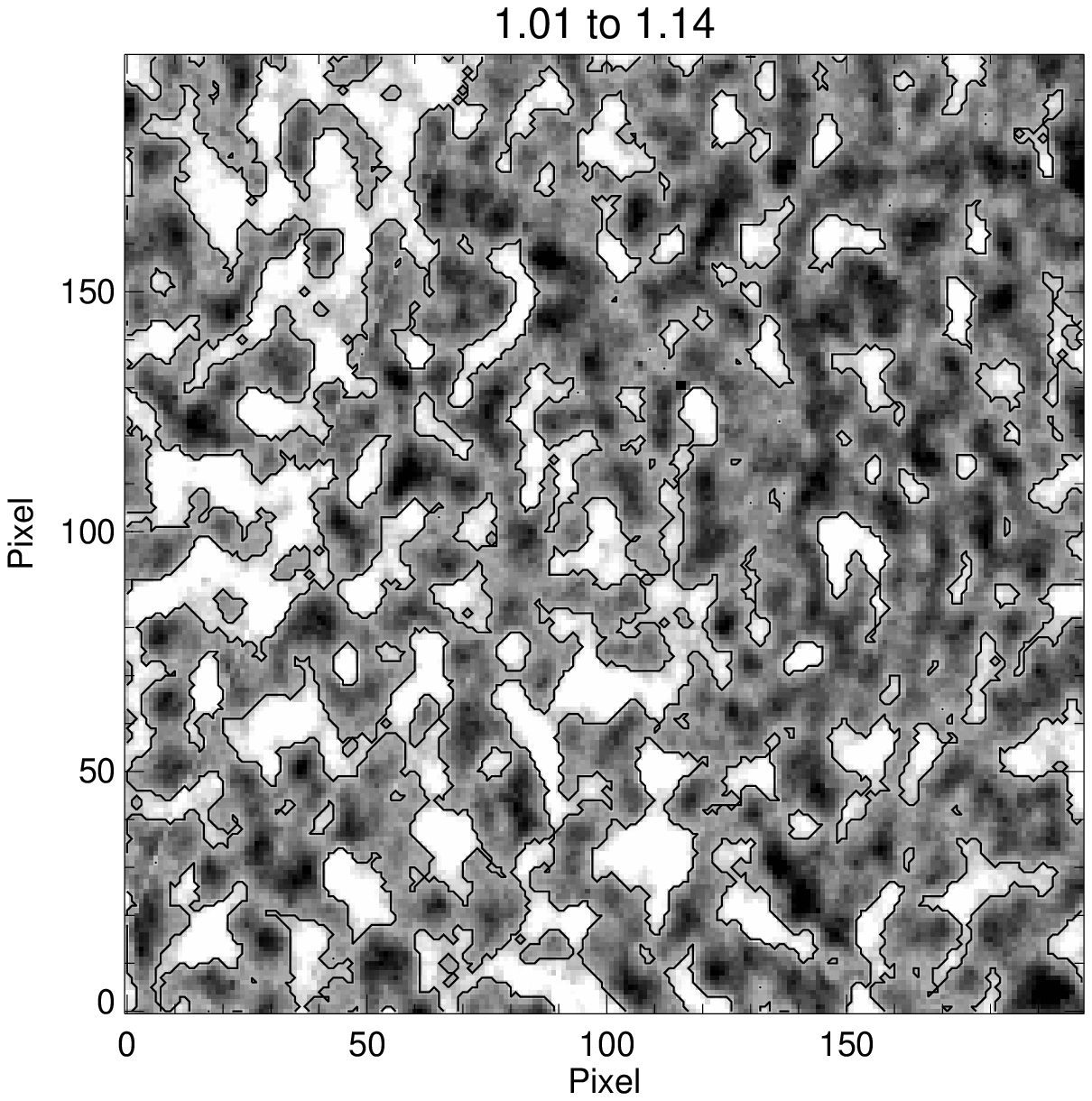} \\
\includegraphics[width=60mm]{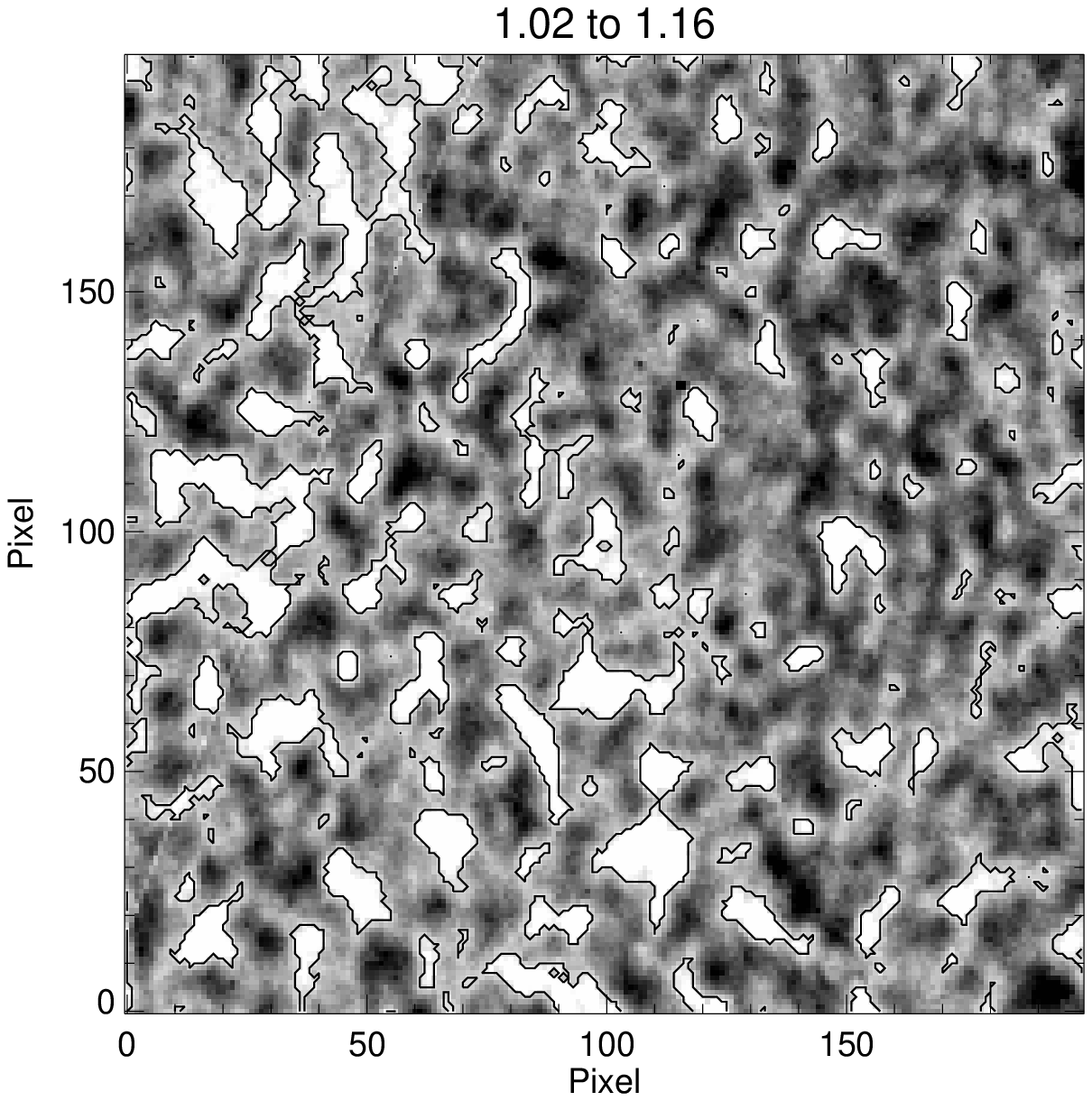} \\
\end{center}
\caption{Same as Fig.~7, but for different values of threshold.}
\label{fig:8}
\end{figure} 

\subsection{Thresholding and identification}

\begin{figure}
\begin{center}
\includegraphics[width=90mm]{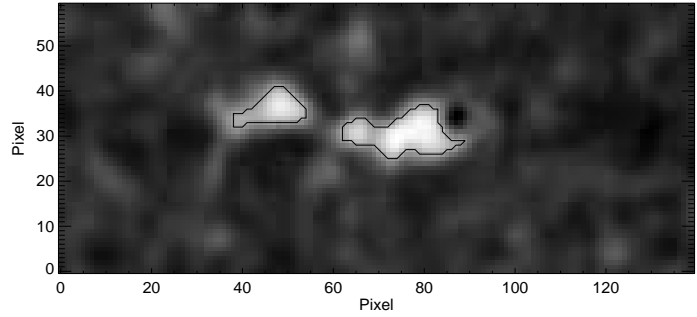}\\
\end{center}
\caption{Contour of the plage region detected by the algorithm is
overlaid upon the image enclosing the plage region.}
\label{fig:9}
\end{figure}

Once the threshold has been decided quantitatively it is an easy task to examine how the values
identify the network elements over the solar disk.  In order to do that we overlaid the 
contours of the 
identified network elements based on the thresholds upon the processed image. For this purpose 
we used a small portion of the image of about 200$\times$200 pixel around the  disk 
center. Figs.~\ref{fig:7} and \ref{fig:8}
show the images extracted from the central portion of the Sun and overlaid with
the contours of the identified network elements with various lower and upper threshold values.
In Fig.~\ref{fig:7} (top and middle rows) contours are obtained by varying the upper 
threshold values and keeping the lower threshold values at 1.002 above the background intensity 
level. The bottom row images show the contour levels that are obtained by varying the lower threshold
levels while keeping the upper threshold levels fixed at 1.14 above the background value.
The contours shown in Fig.~\ref{fig:8} (top row) are obtained by varying the lower threshold values 
and by keeping the upper threshold values at 1.14 above the background level. In the bottom image 
the lower and upper threshold values are kept at 1.02 and 1.16, respectively, above the background value.
From contour maps alone it is difficult to choose an optimum value of upper and 
lower threshold values. It is because most of the images show similar identification. Based 
on the contrast plot (Fig.~\ref{fig:6}) and many such contour images of different days we 
decided to choose 
1.006 as the lower threshold and 1.14 as the upper threshold value. Once the network elements 
have been identified it is easy to isolate the plage regions. The intensity values above
1.14 isolate the plage regions. Fig.~\ref{fig:9} shows the plage region identified using
the threshold value of 1.14 above the background level.

\begin{figure}
\begin{center}
\includegraphics[width=100mm]{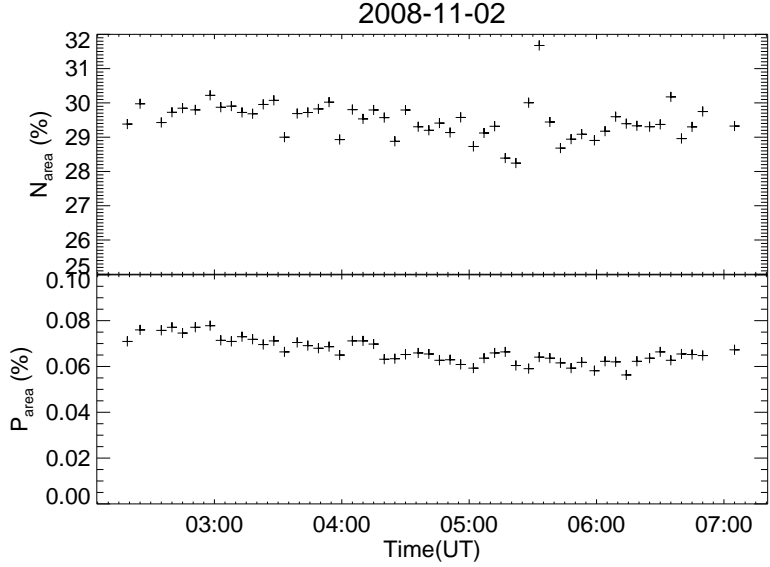}\\
\end{center}
\caption{Network element area (top) and plage area (bottom) normalized to the Sun's visible 
area is plotted in terms of percentage for the data of November 02, 2008.}
\label{fig:10}
\end{figure} 

\subsection{Network element area index}

The network element area index estimates the area occupied by the 
network elements over the solar disk.  It is expressed as a ratio of the
area of the network elements occupied on the Sun to the area of
the visible solar disk. Only the inner 90\% of the solar disk is 
utilized in computing the network element area index. The remaining outer 10\% 
of the disk is not used for the network element area calculation as the 
projection effect is quite large. We have computed the radius of
the Sun for a particular epoch. The area occupied by all the pixels
in contrast range from 1.006 to 1.14 was taken for network element
area index calculation and the resulting network element
area index is converted into percentage using the value of 90\% of visible 
area of the solar disk as described above.

The November 02, 2008 data set has a uniform cadence of Ca-K images and 
available for a long time period of about 5~hrs. Hence, we used the November 02 data
here for the network element area index computation.
Fig.~\ref{fig:10}(top) and (bottom) shows the network element and plage area index for the day
of November 02, 2008. From the plot, it is clear that the network element area index varies
over a day. This variation from morning to evening indicates that the
determination of network element area index is sensitive to the seeing conditions. This is assuming
the fact that the total network element area will not change over the disk within a few hours
of observations.
When the observation starts in the morning, normally the seeing conditions are good at
Kodaikanal. The power to the Ca-K filter is switched ON when the
observations starts to avoid any damage to the filters in the night when no
observer is at the telescope. The passband of the filter takes about 20 min to
reach the central value. The lower Ca-K network element area index values in the 
beginning of the
observations could be the result of small off band of the filter. It could also
be due to the low altitude of the Sun in the early morning and hence large air
column could be disturbing the contrast of the image.  As the altitude of the Sun
increases, the ground heating begins after 5:00~UT. This increases the atmospheric
turbulence and hence seeing becomes poor with time. As this happens,
the quality of the image degrades hence the network boundaries observed in Ca-K image 
becomes diffuse. This  would cause the intensity of the network elements 
below the threshold value of 1.006. 
Hence the poor seeing conditions affect the estimation
of the area of network elements. The variation in the determination of network element area
index due to seeing conditions is marginal at about 1-2\% in the morning hours.
The variation in the area index is moderately large in the noon when the seeing becomes poor by 
about 5$^{\prime\prime}$.
From the plot it is clear that the network elements cover about 30\% of the solar disk.
On November 02, there was a small plage in the field-of-view and its area was about 0.07\% of
the visible solar disk. A small but noticeable decrease in its area was observed over the day.
This could be due to the seeing variations or changes in the plage itself.  
A daily average of network element area index and plage area index for about 3 years (from 
February 2008 to February 2011) has been estimated and 
it is found that the network element occupied about 30\% of the solar disk and the plage
occupied less than 1\% of the solar disk. It is also found that the network element
area index decrease from February 2008 to February 2011 by about 7\%.  Further details about the 
result can be found in Singh et~al. (2012).

\section{Summary}
We have designed, developed and fabricated a Twin Telescope and installed at Kodaikanal
Observatory. The instrument continues to provide data since February 2008 till date. 
This instrument is built to continue the Ca-K observations from Kodaikanal that has a history of
100 years of data obtained earlier in photographic plates. While the digitization of the historic
data is going on, we have planned to develop some of the techniques to automatically identify 
various features observed in different wavelengths. We have applied an intensity threshold 
method for the processed Ca-K images obtained from Twin Telescope
to identify the  network elements and plages. Though,
the method has some limitations,  it is the first step in identifying the features. The main
aim of the paper here is to announce the availability of a new telescope at Kodaikanal for the
solar community and also to provide the information about the method of data calibration and 
reduction. In future, the calibrated data will be available to the researchers via
the world wide web\footnote{http://kso.iiap.res.in:8080/KodaiTwinTelData/TwinTelCatalog.html}  
interface of Indian Institute of Astrophysics. This provision
will be made to provide the current day hourly updates on Ca-K images to the solar community.
In future, the white light images for which the calibration process is currently underway, will
also be made available to the solar community.

\section*{Acknowledgments}
We thank the anonymous referee for his/her fruitful comments that helped us to improve
the presentation in the manuscript.
J.S. is thankful to F. Gabriel and his team for designing,
fabricating and installing the telescope at Kodaikanal, Anbazhagan and K. Ravi
for developing the guiding system for the telescope, S. Muneer for
the initial involvement in the project, F. George, S. Ramamoorthy, P. Loganathan,
P. Michael, P. Devendran, G. Hariharan, Fathima and S. Kamesh for their help to execute
different parts of the project and R. Selvendran and P. Kumaravel
for the daily solar observations using the Twin Telescope. Authors thank Muthu Priyal, 
T.G. Priya and K. Amareswari for calibrating the data and making it available to the users.

\label{lastpage}
\end{document}